\documentclass[twoside,twocolumn,9pt]{article}
\usepackage{extsizes}
\usepackage[super,sort&compress,comma]{natbib} 
\usepackage[version=3]{mhchem}
\usepackage[left=1.5cm, right=1.5cm, top=1.785cm, bottom=2.0cm]{geometry}
\usepackage{balance}
\usepackage{mathptmx}
\usepackage{sectsty}
\usepackage{graphicx} 
\usepackage{lastpage}
\usepackage[format=plain,justification=justified,singlelinecheck=false,font={stretch=1.125,small,sf},labelfont=bf,labelsep=space]{caption}
\usepackage{float}
\usepackage{fancyhdr}
\usepackage{fnpos}
\usepackage[english]{babel}
\addto{\captionsenglish}{%
  
}
\usepackage{array}
\usepackage{droidsans}
\usepackage{charter}
\usepackage[T1]{fontenc}
\usepackage[usenames,dvipsnames]{xcolor}
\usepackage{setspace}
\usepackage[compact]{titlesec}
\usepackage{hyperref}

\usepackage{epstopdf}

\definecolor{cream}{RGB}{222,217,201}

\newcommand{\ju}[1]{\textcolor{black}{#1}}

\begin{document}

\pagestyle{fancy}
\thispagestyle{plain}
\fancypagestyle{plain}{
\renewcommand{\headrulewidth}{0pt}
}

\makeFNbottom
\makeatletter
\renewcommand\LARGE{\@setfontsize\LARGE{15pt}{17}}
\renewcommand\Large{\@setfontsize\Large{12pt}{14}}
\renewcommand\large{\@setfontsize\large{10pt}{12}}
\renewcommand\footnotesize{\@setfontsize\footnotesize{7pt}{10}}
\makeatother

\renewcommand{\thefootnote}{\fnsymbol{footnote}}
\renewcommand\footnoterule{\vspace*{1pt}%
\color{cream}\hrule width 3.5in height 0.4pt \color{black}\vspace*{5pt}} 
\setcounter{secnumdepth}{5}

\makeatletter 
\renewcommand\@biblabel[1]{#1}            
\renewcommand\@makefntext[1]%
{\noindent\makebox[0pt][r]{\@thefnmark\,}#1}
\makeatother 
\renewcommand{\figurename}{\small{Fig.}~}
\sectionfont{\sffamily\Large}
\subsectionfont{\normalsize}
\subsubsectionfont{\bf}
\setstretch{1.125} 
\setlength{\skip\footins}{0.8cm}
\setlength{\footnotesep}{0.25cm}
\setlength{\jot}{10pt}
\titlespacing*{\section}{0pt}{4pt}{4pt}
\titlespacing*{\subsection}{0pt}{15pt}{1pt}

\fancyfoot{}

\fancyhead{}
\renewcommand{\headrulewidth}{0pt} 
\renewcommand{\footrulewidth}{0pt}
\setlength{\arrayrulewidth}{1pt}
\setlength{\columnsep}{6.5mm}
\setlength\bibsep{1pt}

\makeatletter 
\newlength{\figrulesep} 
\setlength{\figrulesep}{0.5\textfloatsep} 

\newcommand{\topfigrule}{\vspace*{-1pt}%
\noindent{\color{cream}\rule[-\figrulesep]{\columnwidth}{1.5pt}} }

\newcommand{\botfigrule}{\vspace*{-2pt}%
\noindent{\color{cream}\rule[\figrulesep]{\columnwidth}{1.5pt}} }

\newcommand{\dblfigrule}{\vspace*{-1pt}%
\noindent{\color{cream}\rule[-\figrulesep]{\textwidth}{1.5pt}} }

\makeatother

\twocolumn[
  \begin{@twocolumnfalse}

\vspace{1em}
\sffamily
\begin{tabular}{m{4.5cm} p{13.5cm} }

 & \noindent\LARGE{\textbf{Active nematics with deformable particles$^\dag$}} \\
\vspace{0.3cm} & \vspace{0.3cm} \\

& \noindent\large{Ioannis Hadjifrangiskou, Liam J. Ruske, and Julia M. Yeomans} \\

& \hspace{-10cm}\noindent\normalsize{

The hydrodynamic theory of active nematics has been often used to describe the spatio-temporal dynamics of cell flows and motile topological defects within soft confluent tissues. Those theories, however, often rely on the assumption that tissues consist of cells with a fixed, anisotropic shape and do not resolve dynamical cell shape changes due to \ju{flow gradients}. In this paper we extend the continuum theory of active nematics to include cell shape deformability. We find that circular cells in tissues must generate sufficient active stress to overcome an elastic barrier to deforming their shape in order to drive tissue-scale flows. Above this threshold the systems enter a dynamical steady-state with regions of elongated cells and strong flows coexisting with quiescent regions of isotropic cells.
} \\

\end{tabular}

 \end{@twocolumnfalse} \vspace{0.6cm}

  ]

\renewcommand*\rmdefault{bch}\normalfont\upshape
\rmfamily
\section*{}
\vspace{-1cm}


\footnotetext{The Rudolf Peierls Centre for Theoretical Physics, Beecroft Building, Parks Road,
Oxford, OX1 3PU, UK. E-mail: ioannis.hadjifrangiskou@physics.ox.ac.uk, liam.ruske@physics.ox.ac.uk, julia.yeomans@physics.ox.ac.uk  }




\section{\label{sec:intro}Introduction}
Living systems are inherently active as they use chemical energy from their surroundings to do work. It is becoming increasingly clear that many features seen in cell layers and tissues can be understood using the continuum theory of active nematics, which describes hydrodynamic interactions between active anisotropic particles. Examples include biofilm initiation, topological defects in cell monolayers  and epithelial expansion.\ju{\cite{blanch2017hydrodynamic,blanch2018turbulent,saw2018biological,yaman2019emergence, pearce2019flow,duclos2017topological,pearce2021orientational,saw2017topological,copenhagen2021topological}}

While the theory of active nematics predicts many qualitative features such as short-range orientational order, active turbulence and motile topological defects,\ju{\cite{ardavseva2022topological, mueller2019emergence,thampi2016active,alert2022active,doostmohammadi2016defect}} it is based on liquid crystal hydrodynamics which assumes that the active particles are nematogens with a fixed aspect ratio.\cite{de1993physics} While this assumption may be a reasonable description for systems consisting of rod-shaped cells such as fibroblasts or  {\it Escherichia coli},\cite{dell2018growing} many particles, such as MDCK cells or soft colloids, can undergo large shape changes and their aspect ratio can vary significantly due to active forces.\cite{vlassopoulos2014tunable, lecuit2007cell, heisenberg2013forces}
Simulations have shown that intercellular stresses in monolayers are enhanced by cell deformation, creating a positive feedback loop that affects the collective behaviour of the layer.\cite{balasubramaniam2021investigating}

In this paper we extend the continuum equations which describe active nematic fluids by considering not only the magnitude and direction of the nematic order, but also the shape dynamics of the underlying particles. We consider an equation of motion for the aspect ratio of the deformable nematogens which incorporates flow-driven stretching and compression of particles, as well as elastic restoring forces. The aspect ratio of particles in turn affects the orientational dynamics by modifying thermodynamic interactions between particles and the way they align in shear flows. 

In section \ref{sec:theory} we present the equations of motion for particle shapes, characterised by the aspect ratio $\omega$, the nematic order tensor $\mathbf{Q}$ of the particle orientational distribution, and the associated velocity field $\mathbf{u}$.  
The following sections, \ref{sec:inst} and \ref{sec:num_invest}, are analytical and numerical investigations of systems consisting of elastic, active particles which are isotropic in the absence of flow. We report a shape instability in which extensile active stress generated by particles drives the formation of regions with highly anisotropic particles with nematic order. 
 Finally, in section \ref{sec:discussion} we summarize our results. 
\section{\label{sec:theory}Equations of motion}
\begin{figure}[htp]
    \centering
    \includegraphics[width = 8cm]{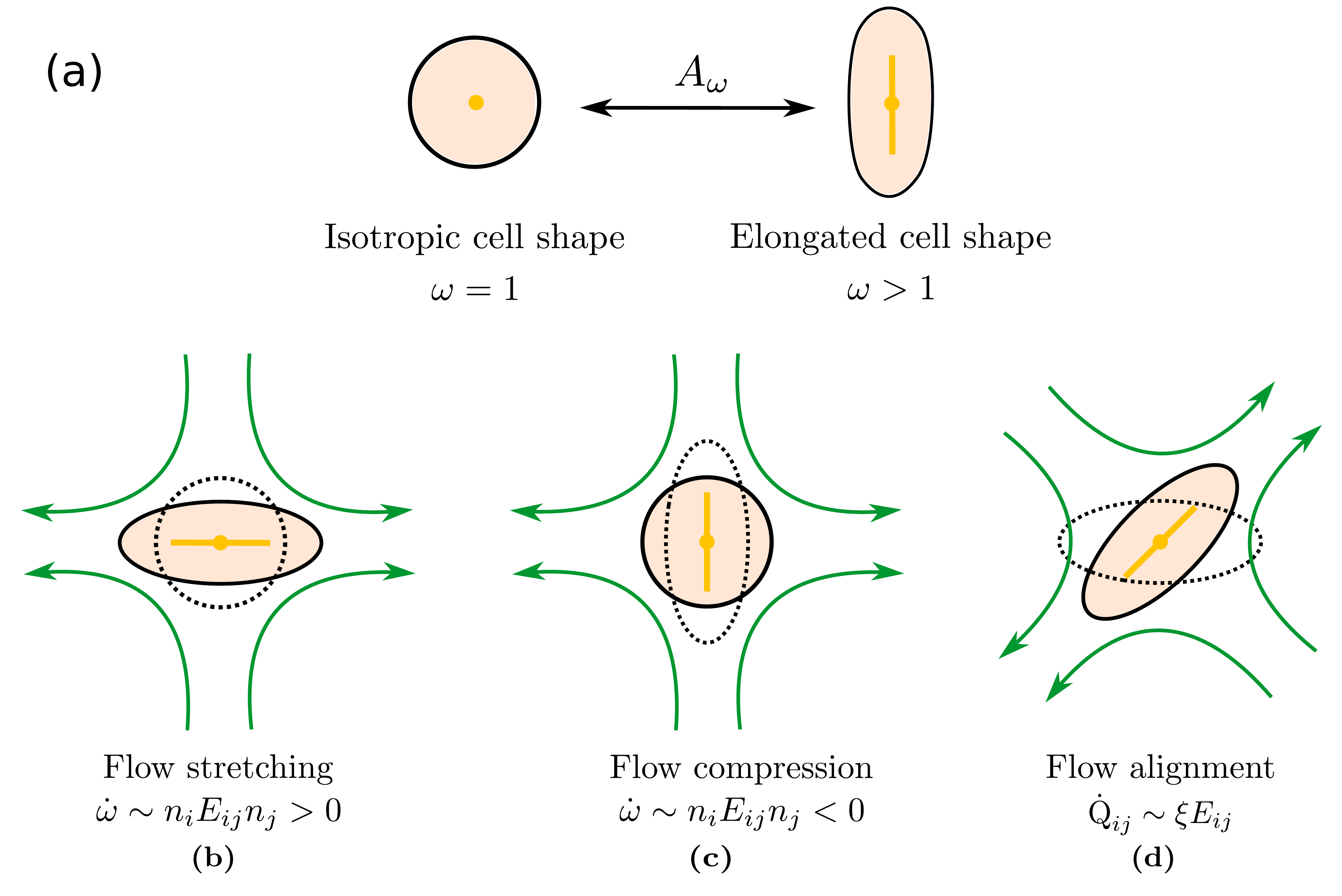}
    \caption{Particle deformations driven by (a) free energy minimisation, (b) flow stretching, (c) flow compression, (d) flow alignment.}
    \label{fig:cell_deformation}
\end{figure}

We introduce coarse-grained equations of motion to describe the hydrodynamics of elastic, nematic particles. To this end we approximate the shape of particles as ellipsoidal, and define a field $\omega > 1$ to describe their local aspect ratio, Fig.~\ref{fig:cell_deformation} (a). 
The time evolution of the shape distribution follows \cite{bilby1977finite, gao2011rheology, gao2013dynamics} 
\begin{align}\label{omega:time_evo}
    \left( \partial_{t} + \mathbf{u} \cdot \mathbf{\nabla} \right)\omega - 2\omega E_{\parallel} = -\Gamma_{\omega}\dfrac{\delta \mathcal{F} }{\delta \omega}. 
\end{align}
The first term in Eq.~(\ref{omega:time_evo}) describes advection by a macroscopic flow field $\mathbf{u}$, while the second term accounts for shape changes due to flow gradients. $E_{\parallel}=n_{i}E_{ij}n_{j}$ is the projection of the strain rate tensor $E_{ij}=(\partial_i u_j + \partial_j u_i)/2$  along the long axis of the particles defined by the unit vector $\mathbf{n}$.
Depending on the particle orientation with respect to the extensional flow axis, particles \ju{are} either stretched or compressed by flows (Fig.~\ref{fig:cell_deformation} (b), (c)). The final term in Eq.~(\ref{omega:time_evo}) models particle shape elasticity, where $\Gamma_{\omega}$ controls the relaxation towards the minimum of a free energy $\cal{F}$ as defined below.

Nematic ordering of anisotropic particles is described by a tensorial order parameter, $Q_{ij}= 3 S/2 \: \left( n_i n_j - \delta_{ij}/3 \right)$, where $S$ quantifies the magnitude of the nematic alignment, and we restrict the directors to lie within a plane so that $n_{z} = 0$ and $u_{z} = 0$. The time evolution of $\mathbf{Q}$ follows the Beris-Edwards equation,\cite{beris1994thermodynamics}
\begin{equation}
	\left( \partial_{t} + \mathbf{u} \cdot \mathbf{\nabla} \right) \mathbf{Q} - \mathcal{W} = \Gamma_{LC} \mathbf{H} , \:
	\label{EoM_Q}
\end{equation}
where $\Gamma_{LC}$ controls the relaxation towards an equilibrium quantified by the molecular field $\mathbf{H} = -\delta \mathcal{F}/\delta \mathbf{Q} + (\mathbf{I}/3) \text{Tr}( \delta \mathcal{F} / \delta \mathbf{Q})$. Nematogens are not only advected by the fluid, but also rotated by gradients in the flow field, which gives rise to the the co-rotational term \citep{beris1994thermodynamics} 
\begin{align} \label{corot}
	\mathcal{W}_{ij} = &\left( \xi E_{ik}+\Omega_{ik} \right) \left( Q_{kj} + \frac{\delta_{kj}}{3} \right) + \left( Q_{ik} + \frac{\delta_{ik}}{3} \right) \left( \xi E_{kj}-\Omega_{kj} \right) \nonumber \\ 
    &- 2\xi \left( Q_{ij} + \frac{\delta_{ij}}{3} \right) Q_{kl} W_{lk} ,
\end{align}
where $\Omega_{ij}=(\partial_j u_i - \partial_i u_j)/2$ is the vorticity tensor, the antisymmetric part of the velocity gradient tensor $W_{ij}=\partial_i u_j$. The flow-alignment parameter $\xi$ quantifies the rotation of anisotropic particles in shear flows (Fig.~\ref{fig:cell_deformation}(d)). Its value depends on the shape of the particles and it is zero for isotropic particles.\cite{jeffery1922motion} Therefore we assume $\xi = \xi_{0}(\omega - 1)$ where $\xi_{0}$ sets the flow-alignment scale. 
\begin{figure*}
    \centering
    \includegraphics[width = 12cm]{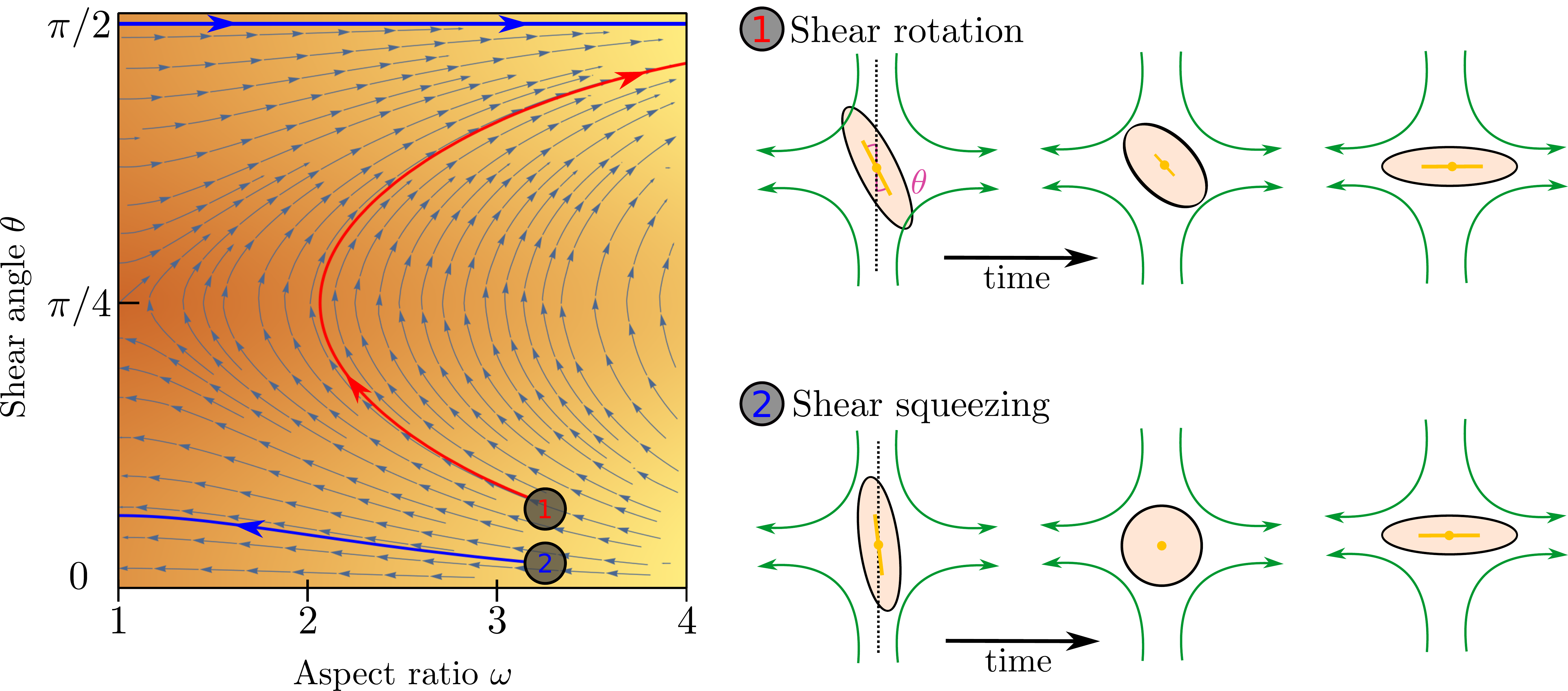}
    \caption{Left: Phase space trajectories of the evolution of $\omega$ subject to an extensional flow at an angle $\theta$ to the director. Right: Rotation and squeezing of a particle in an extensional flow.} 
    \label{fig:cell_phasedia}
\end{figure*}

The total free energy of the system, $\mathcal{F} = \int (f_{\omega} + f_{LC})\: d\mathbf{V}$, consists of two contributions.
The first contribution is an elastic energy associated with particle shape deformations, 
\begin{align}
	f_{\omega} = A_{\omega}\frac{1}{2}(\omega - \omega_{0})^{2} + A_{\omega}^{\star}\frac{1}{4}(\omega - \omega_{0})^{4} \: ,
	\label{fdef}
\end{align}
where $\omega_0$ is the particle aspect ratio at equilibrium and $A_{\omega}$, $A_{\omega}^{\star}$ are elastic deformation parameters. The second represents the liquid crystal free energy which emerges from particle-particle interactions,
\begin{align}
    f_{LC} = &\dfrac{A_{LC}}{2}\left(S_0^{2} \: \left(\frac{\omega - 1}{\omega}\right)^2 - \dfrac{2}{3}\mbox{tr}(\boldsymbol{Q}^{2})\right)^{2} + \frac{1}{2} K_{LC} \left( \mathbf{\nabla} \mathbf{Q} \right)^{2},
\end{align}
where $K_{LC}$ penalizes distortions in the director field in the one-elastic-constant approximation and $A_{LC}$ sets the bulk-energy scale. 
Highly anisotropic particles tend to align nematically due to excluded volume effects whereas there is no alignment interaction between circular particles. This is reflected in the free energy which has a minimum at a reference value, $S=S_0$, for $\omega \rightarrow \infty$ while favouring $S=0$ for $\omega=1$. 

Finally, the velocity field $\mathbf{u}$ of the fluid obeys the incompressible Navier-Stokes equations,
 \begin{equation}
 \nabla \cdot \mathbf{u} = 0 \: ,
 \label{EoM_compress}
 \end{equation}
\begin{equation}
	\rho \left( \partial_{t} + \mathbf{u} \cdot \mathbf{\nabla} \right) \mathbf{u} = \mathbf{\nabla} \cdot \Pi , \:
	\label{EoM_fluid}
\end{equation}
where $\rho$ is the fluid density and the stress tensor $\Pi = \Pi^{passive} + \Pi^{active}$ includes passive and active contributions. The passive stress tensor consists of viscous dissipation and elastic stress arising from liquid crystal hydrodynamics,\cite{marenduzzo2007hydrodynamics}

\begin{equation}
 \Pi^{viscous} = 2 \eta \mathbf{E} \:,
\end{equation}
\begin{align}
 \Pi^{elastic} =& -p \mathbf{I} - \xi [ \mathbf{H} \Tilde{\mathbf{Q}} + \Tilde{\mathbf{Q}} \mathbf{H} - 2 \Tilde{\mathbf{Q}}  \mbox{tr}(\mathbf{Q} \mathbf{H}) ] + \mathbf{Q} \mathbf{H} \nonumber \\ 
 &- \mathbf{H} \mathbf{Q} - \mathbf{\nabla} \mathbf{Q} \left( \frac{\partial f_{LC}}{\partial (\mathbf{\nabla} \mathbf{Q})} \right),   
 \end{align}
where $\eta$ is the viscosity, $p$ is the bulk pressure and $\Tilde{\mathbf{Q}} = \left(\mathbf{Q}+\frac{1}{3} \mathbf{I}\right)$. Active dipolar forces produced by individual particles give rise to an active stress,\cite{simha2002hydrodynamic}
\begin{equation}
	\Pi^{active} = - \zeta \mathbf{Q} \: ,
\end{equation}
where $\zeta$ quantifies the magnitude of active forces. $\zeta>0$, corresponds to extensile activity, where fluid is pushed outwards from the particles along their direction of elongation, and pulled inwards along the perpendicular axis. $\zeta<0$ corresponds to contractile activity, where the flow direction is reversed. We consider extensile systems in what follows. 

We use a hybrid lattice Boltzmann--finite difference method  to solve the equations of motion (\ref{omega:time_evo}, \ref{EoM_Q}, \ref{EoM_compress}, \ref{EoM_fluid}).\cite{marenduzzo2007steady} The lattice size is $200 \times 200$ with a lattice spacing $\Delta x = 1$, LB timestep $\Delta t = 1$ and periodic boundary conditions. Unless stated otherwise, the simulation parameters are $\Gamma_{LC} = 0.1, A_{LC} = 0.1, S_{0} = 1, K_{LC} = 0.015, \xi_{0} = 0, \Gamma_{\omega} = 0.01, A_{\omega} = 0.04, A_{\omega}^{\star} = 0.003, \omega_{0} = 1, \zeta = 0.001, \rho = 1, \eta = 2/3$. \ju{The relevant inverse time-scales are as follows: $\Gamma_{\omega}A_{\omega}$, the rate of relaxation of the particles to their preferred shape, $\Gamma_{LC}A_{LC}$, the rate of relaxation of the order parameter towards the free energy minimum, and $\zeta/\eta$, the rate of active stresses injected into the system. In the absence of flow gradients, the regime where $\Gamma_{\omega}A_{\omega} \gg \Gamma_{LC}A_{LC}$ has nematic alignment responding slowly to changes in shape, creating a time delay associated with how particles will align after their shape changes. In the inverse case, nematic alignment responds quickly to shape changes, which is the regime we are considering. The relevant length-scales are the active length-scale $L_{act} \sim \sqrt{K_{LC}/\zeta}$ and a passive length-scale $L_{nem} \sim \sqrt{K_{LC}/A_{\omega}}$, which sets the scale of shape gradients.}
Simulations are initialised with random noise in $\omega$ around a finite value of $\omega = 1.3$ and $S = \left(\omega - 1\right)/\omega$. We use a runtime of $2 \times 10^{5}$ LB timesteps. Initial conditions become irrelevant after a runtime of $\approx 10^5$ timesteps. If $\omega$ falls below unity, we rotate the director field to point towards the extensile flow axis. This is physically motivated by considering a particle that is instantaneously circular; any subsequent elongation will occur along the local extensile axis of the active flow field. 

\section{Results}
To illustrate the coupled time evolution of particle orientation $\mathbf{n}$ and aspect ratio $\omega$ we consider a passive system in an external extensional flow and solve Eq.~\eqref{EoM_Q} for the director field in the Ericksen-Leslie limit (constant $S$) \cite{beris1994thermodynamics} for a finite flow-aligning parameter which scales linearly with the aspect ratio, coupled to Eq.~\eqref{omega:time_evo} with $\Gamma_{\omega} = 0$. Depending on the position in phase space, there are two distinct re-orientation mechanisms: If the angle $\theta$ between the compressive flow axis and particle orientation is sufficiently large, particles undergo rotation where their shape stays anisotropic and their orientation rotates until it aligns with the extensional flow axis  (red trajectory in Fig.~\ref{fig:cell_phasedia}).
If $\theta$ is small, however, elongated particles cannot rotate fast enough and get squeezed by the flow until they become isotropic, $\omega=1$. Subsequently their orientation aligns along the extensional flow axis, $Q_{ij} \sim E_{ij}$, and their aspect ratio increases again  (blue trajectory in Fig.~\ref{fig:cell_phasedia}). 

\subsection{\label{sec:inst} Linear Stability Analysis}
To investigate the behaviour of an active system consisting of nearly circular particles, we perform a stability analysis around a quiescent isotropic system by adding small perturbations to the aspect ratio of particles, $\omega = 1 + \delta \omega$. The shape perturbation $\delta \omega \sim \exp{(i \mathbf{q} \cdot \mathbf{r} + \lambda t)}$ is applied along a wavevector $\mathbf{q} = q\left(\cos{\theta}, \sin{\theta},0\right)$ and we assume particles deform along their long axis. The flow fields 
which result from activity give rise to the dispersion relation,
\begin{align}\label{dispersion_relation}
    &q^{2}\left\{\left(\eta +  \dfrac{3}{2}\Gamma_{\omega}K_{LC}\rho\right)\lambda + \Gamma_{\omega}A_{\omega}\eta - \dfrac{3}{4}\zeta \sin^{2}{2\theta}\right\} \notag\\
    &+ \dfrac{3}{2}\Gamma_{\omega}K_{LC}\eta q^{4} + \rho \lambda^{2} + \Gamma_{\omega}A_{\omega}\rho\lambda = 0 \: .
\end{align}

\begin{figure}[htp]
    \centering
    \hspace{-0.8cm}
    \centerline{\includegraphics[width = 8cm, height = 7cm]{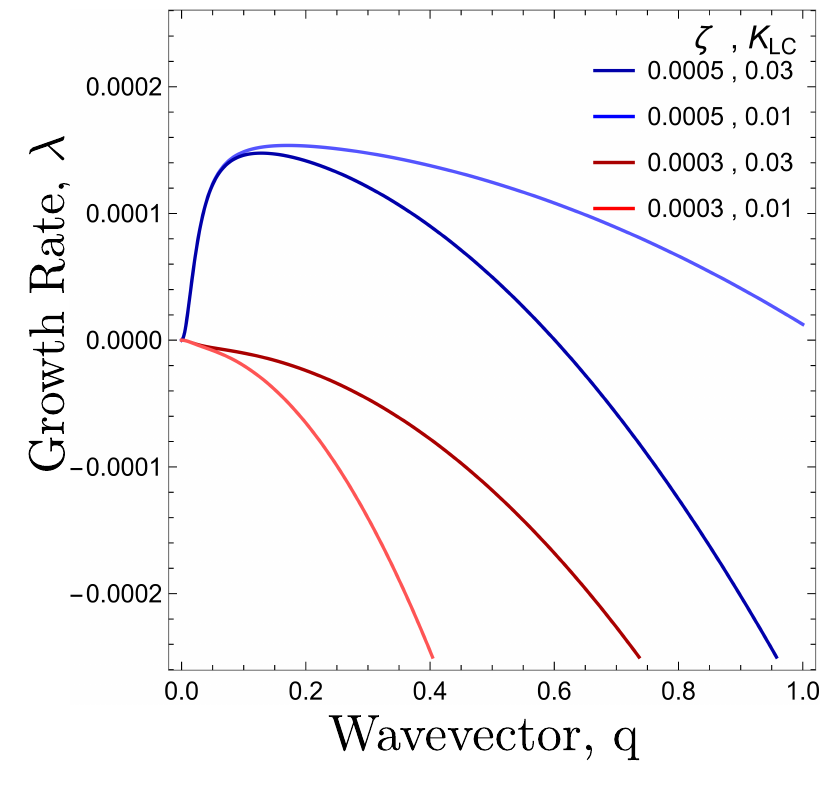}}
    \hspace{-0.8cm}\caption{Dispersion relation of $\lambda(q)$ from Eq.~\eqref{dispersion_relation} plotted for $\theta = \pi/4$ for different values of \{${\zeta, K_{LC}}$\}. Red/Blue curves correspond to activity values smaller/greater than the threshold activity, $\zeta_{c}$ respectively.}
    \label{fig:disp_relation}
\end{figure}
\begin{figure*}[htp]
    \centering
    \includegraphics[width = 16cm]{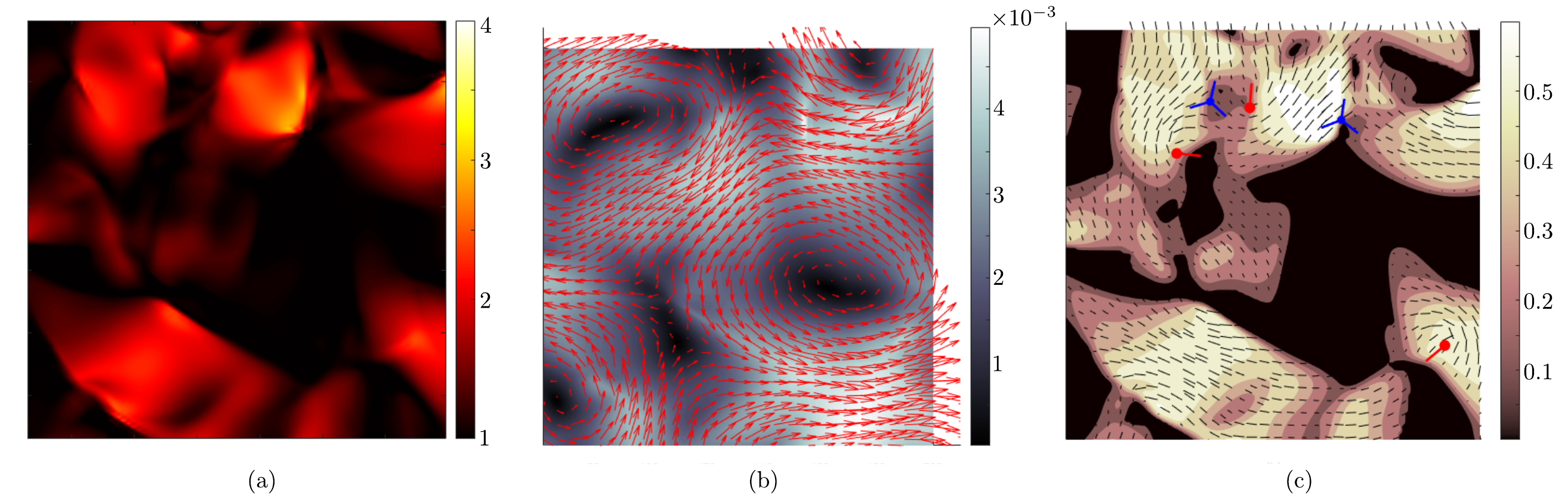}
    \caption{Fully developed active turbulence with deformable particles. (a) Aspect ratio, $\omega$ of particles. (b) Velocity field, $\mathbf{u}$. Colour scale indicates the magnitude of the velocity. (c) Nematic director field $\mathbf{n}$ and scalar order parameter $S$. Colour scale indicates the magnitude of $S$. Red/blue symbols correspond to the core of $\pm 1/2$ defects respectively. $+1/2$ defects are oriented such that the tail is along the line shown.  }
    \label{fig:instability}
\end{figure*}
For a solution with $\lambda > 0$ to exist for some value of $q$, we require the coefficient of the quadratic term to be negative. Thus the critical value of the activity below which an isotropic system will be stable is given by:
\begin{align}
    \zeta_{c} = \dfrac{4\Gamma_{\omega}A_{\omega}\eta}{3\sin^{2}{2\theta}}.
\end{align}
$\zeta_{c}$ is clearly minimised when $\theta = \pi/4$, (Fig.~\ref{fig:disp_relation}). At this angle, the parallel and perpendicular components of the flow work in tandem to create the most efficient elongation of the particles, see Appendix \ref{append:C}. 
Previously, activity was found to create nematic order due to flow aligning effects as long as $\xi \zeta > 0$.\cite{santhosh2020activity} Here we provide a mechanism where activity may induce nematic order without the need for flow alignment. Starting from an isotropic system of circular particles, activity causes elongation and then thermodynamic forces described by the free energy may allow for nematic order to be established. 
\\
The competition becomes apparent if we explicitly write down the wavevector below which the instability grows:
\begin{align}
    q_{0}^{2} = \dfrac{3\zeta \sin^{2}{2\theta}- 4 \Gamma_{\omega}A_{\omega}\eta}{6K_{LC}\Gamma_{\omega}\eta}.
\end{align}
This depends on a balance between the active length-scale, $L_{act} \sim \sqrt{K_{LC}/ \zeta}$, and the passive length-scale,  $ L_{nem} \sim \sqrt{K_{LC}/A_{\omega}}$.
\subsection{\label{sec:num_invest}Active turbulence} 
\begin{figure}[htp]
    \centering
    \centerline{\includegraphics[width = 8cm, height = 3.5cm]{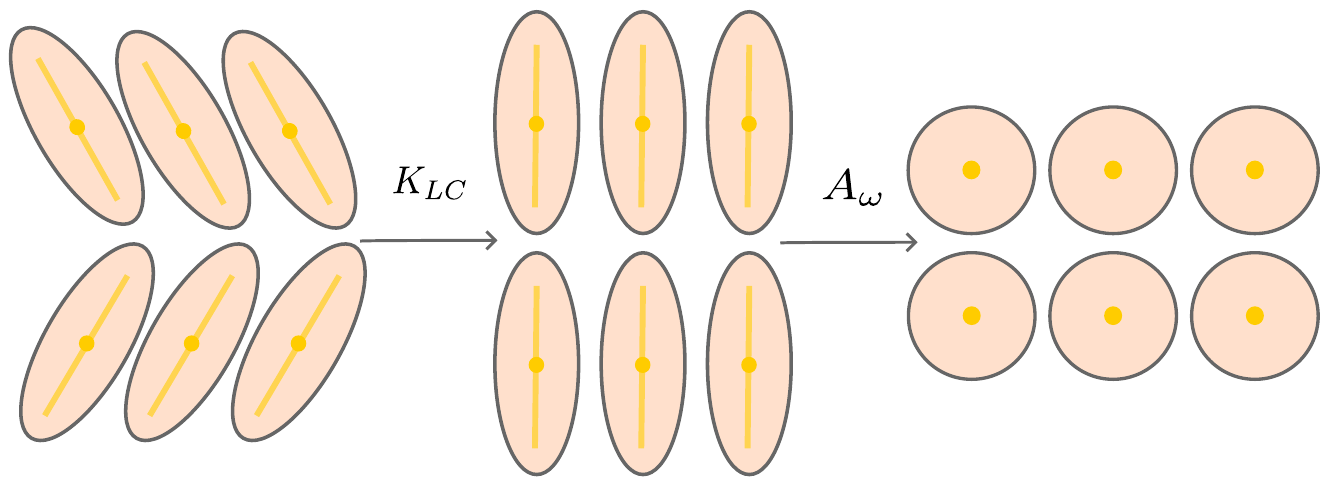}}
    \caption{Creation and contraction of nematic domains. Nematic elasticity smooths out gradients in the director field, resulting in a domain of aligned directors. The active force vanishes and particles contract due to the elastic free energy cost.}
    \label{fig:cell_alignment_contraction}
\end{figure}

\begin{figure}[H]
    \centering
    \vspace*{-0.3cm}
    \centerline{\includegraphics[width = 9cm, height = 7cm]{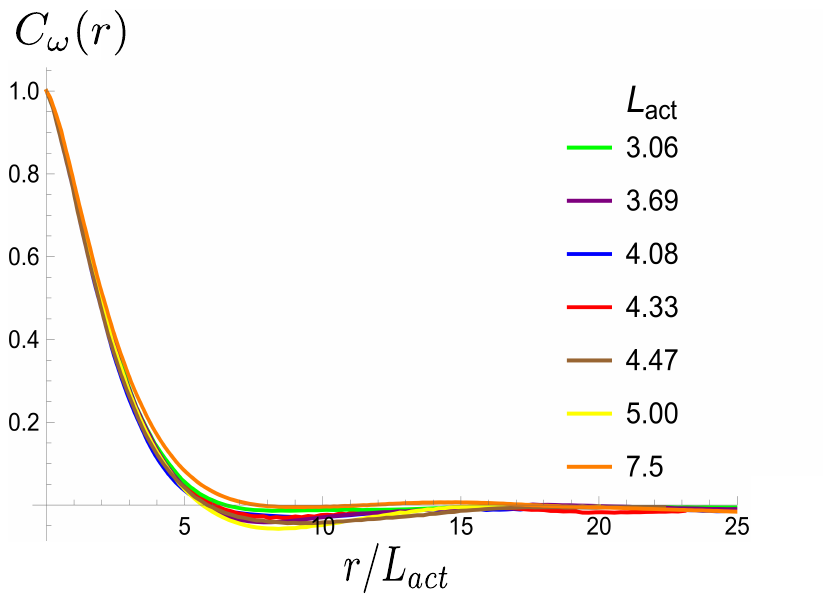}}
    \caption{Aspect ratio correlation, $C_{\omega}$ as a function of distance, scaled by the active length-scale, $L_{act} = \sqrt{K_{LC}/\zeta}$. Different coloured curves correspond to differing values of $L_{act}$, generated by varying $\zeta$ and $K_{LC}$ independently in systems exhibiting fully developed active turbulence. }
    \label{fig:corr_curves}
\end{figure}

For values of the extensile activity above which flows are maintained the system enters an active turbulent regime.  Fig.~\ref{fig:instability} shows the dynamical steady state corresponding to fully developed active turbulence in the system of deformable particles. At any given point in time the system is characterised by distinct regions of elongated particles separated by areas where there is little or no particle extension. There is strong nematic ordering in the regions where the particles are extended, and it is possible to identify the motile topological defects characteristic of active turbulence. 
In active turbulence with particles of fixed length, nematic domains are broken up by the active instability and by the passage of topological defects. Here there is an additional mechanism which results in regions of circular particles. Within a nematic domain the gradients in $\mathbf{Q}$ are small, and thus active forces are diminished. Consequently, the magnitude of $E_{\parallel}$ drops, and it is no longer able to overcome the free energy cost of deformation, Eq.~(\ref{fdef}). The particles within these domains will therefore begin to contract. Fig.~\ref{fig:cell_alignment_contraction} shows an idealised sketch of this process.
\\
To investigate how physical parameters affect the dynamical state of the system of deformable particles, we measure the time and space averaged aspect ratio, $\langle \omega \rangle$ as the activity parameter $\zeta$, the elastic deformation parameter $A_{\omega}$, the elastic constant, $K_{LC}$ and the flow alignment parameter, $\xi_{0}$ are varied. In all cases, activity was high enough to keep the system in active turbulence.
\begin{figure}
    \centering
    \hspace*{-0.1cm}\includegraphics[width = 9cm, height = 9.0cm]{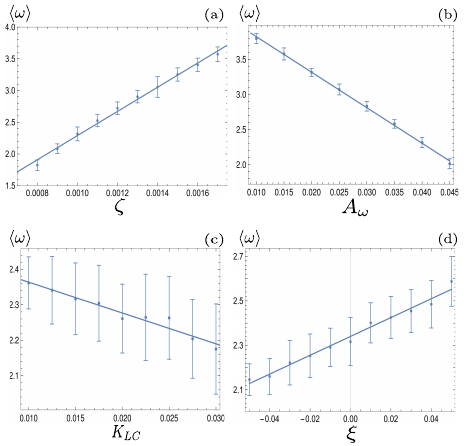}
    \caption{Average aspect ratio as a function of (a) the activity, $\zeta$, (b) elastic deformation parameter, $A_{\omega}$, (c) the elastic constant, $K_{LC}$  and (d) the flow alignment parameter $\xi_{0}$. A linear relationship using a least-squares fit has been plotted in each case. 
       Error bars are given by the standard deviation of the spatial average of the aspect ratio over $10^{5}$ timesteps. }
    \label{fig:omega_plots}
\end{figure}
Fig.~\ref{fig:omega_plots} (a) shows that $\langle \omega \rangle$ increases linearly with the activity for extensile systems. In active turbulence where the evolution of the flow field is dominated by driving from the active stress, the magnitude of the velocity, and consequently the strain rate, increase with the activity parameter, resulting in a higher degree of elongation. 
Conversely, Fig.~\ref{fig:omega_plots} (b) shows that $\langle \omega \rangle$ decreases linearly with $A_{\omega}$. As the elastic energy cost of deforming particles increases, the average aspect ratio will decrease.  
Next, as $K_{LC}$ is increased, gradients in the director field are smoothed out, reducing the overall active stress and consequently reducing elongation. This effect is relatively small compared to the other parameters, as Fig.~\ref{fig:omega_plots} (c) shows. Finally, $\langle \omega \rangle$ also increases linearly with the value of the flow aligning parameter, $\xi_{0}$, shown in Fig.~\ref{fig:omega_plots} (d). As $\xi_{0}$ increases, particles will align more efficiently with the extensional flow axis, thus increasing their elongation.
\\
Next, we look at any emergent length-scales in the system. As we have already shown, in a system of fully developed active turbulence, distinct domains of elongated, anisotropic particles coexist with domains of isotropic particles. To quantify the size of these domains, we calculate the normalised correlation function defined as \ju{$C_{\omega}\left(r\right) = \langle\left(\omega\left(0, t\right) - \Bar{\omega}\left(t\right)\right)\left(\omega\left(\mathbf{r}, t\right) - \Bar{\omega}\left(t\right)\right)\rangle/\langle\left(\omega\left(0,t\right) - \Bar{\omega}\left(t\right)\right)^{2}\rangle$ where $\Bar{\omega}$ is the spatial average of the aspect ratio at each point in time and $\langle \cdot \rangle$ denotes a space and time average over all points separated by a distance $r$.} Fig.~\ref{fig:corr_curves} shows how $C_{\omega}$ varies with distance, scaled by the active length-scale, $L_{act} \sim \sqrt{K_{LC}/\zeta}$ for different choices of the activity and elasticity parameters $\zeta$ and $K_{LC}$. The collapse of the data to a single curve shows that the size of the domains is governed by $L_{act}$, as in active turbulence with fixed length nematogens, reflecting that flow gradients, which occur on this length scale are responsible for the correlations in the aspect ratio of particles.
We also measure the density of topological defects throughout the system as the activity $\zeta$, the elastic constant $K_{LC}$ and the cell elasticity $A_{\omega}$ 
are varied. \ju{To obtain this data we track defects \cite{blow2014biphasic,vromans2016orientational} and also set a cut-off value $\omega = 1.1$ below which topological defects cannot be observed. }

\begin{figure}[H]
        \centering
        \includegraphics[width=9cm, height = 7.0cm]{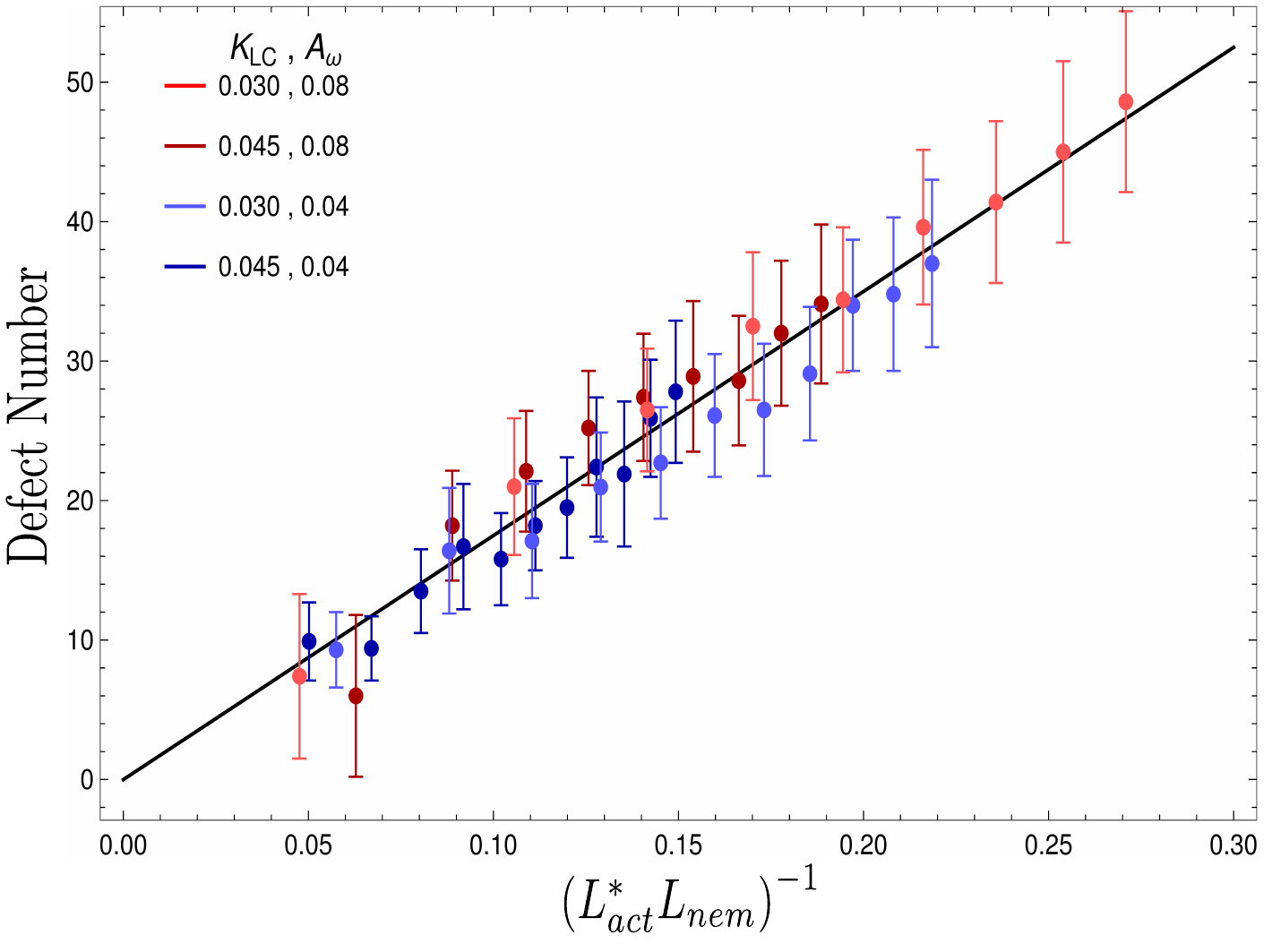} 
        \caption{Average number of topological defects as a function of the inverse modified active length-scale, $L^{*}_{act}$ multiplied by $L_{nem}$. The plotted line shows a least-squares fit through all the data points. Data points were collected by varying the activity parameter, $\zeta$ for different realisations of the parameters $\{K_{LC}, A_{\omega}\}$. Error bars are given by the standard deviation of the spatial average of the defect number over $10^{5}$ timesteps.} 
        \label{fig:def_number}
\end{figure}

As Fig.~\ref{fig:def_number} shows, the defect density scales as $(L^{*}_{act}L_{nem} )^{-1}$ where the modified active length-scale is defined as $L_{act}^{*} \sim \sqrt{K_{LC}/(\zeta - \zeta^{*})}$, where $\zeta^{*}$ is the threshold activity required to keep the system in fully developed active turbulence - this is taken to be the activity value below which circular cells dominate the system and consequently, topological defects cannot exist. An estimate for this threshold activity is given below.
This is an empirical result and is a fundamentally different scaling than fixed length active nematics where the defect density scales with the inverse active length-scale squared, $L_{act}^{-2} = \zeta/K_{LC}$.\cite{giomi2015geometry} Indeed it seems very reasonable that other length-scales should be involved. One clear difference from the usual active nematic models is that the active stresses depend on the local magnitude of the order parameter and therefore vary across the system, particularly at interfaces, and this will influence the defect count. 
\ju{A plausible explanation for this scaling is as follows: $L^{*}_{act}$ plays a similar role to $L_{act}$ in traditional active nematics, with a constant offset in the activity due to a minimum threshold required to achieve active turbulence. Secondly, the deformation energy $A_{\omega}$ will tend to reduce the magnitude of the aspect ratio, $\omega$, and therefore the order parameter, $S$. In turn, this will reduce the effectiveness of the nematic elasticity which depends on $S$. There is then a reduced, effective elastic constant $\sim K_{LC}/A_{\omega}$ so the defect density scales with $L_{nem}^{-1} \sim \sqrt{A_{\omega}/K_{LC}}$. }

An estimate for the threshold activity $\zeta^{\ast}$ may be obtained by using \ju{the} velocity field around $+1/2$ defects \cite{giomi2014defect} to calculate the strain rate, $E_{\parallel}^{+}$. For a director given by $\mathbf{n} = \left(\cos{\theta/2}, \sin{\theta/2},0\right)$ and assuming $S = \left(\omega-1\right)/\omega$ outside the core, the strain rate is given in polar coordinates by
\begin{align}
    &E_{\parallel}^{+}(r, \theta) = \dfrac{\omega-1}{\omega}\dfrac{\zeta}{16\eta}\left(\cos{2\theta} + 3\right) .\label{pos_strn} 
\end{align}
By substituting \eqref{pos_strn} into Eq.~\eqref{omega:time_evo}, neglecting advection and quartic contributions in the free energy, a simple calculation yields a lower bound to the activity required for particles to elongate along the axis of symmetry of $+1/2$ defects:

\begin{align}
    \zeta^{*} = 2\Gamma_{\omega}A_{\omega}\eta.
\end{align}
This estimate ignores elasticity: it is lower than the numerical values by between 10\% for low values of the elastic constant and 50\% for high values of the elastic constant.
~\\
%
\section{\label{sec:discussion}Discussion}
We have extended the continuum theory of active nematics to allow for deformable particles. This was achieved by introducing an equation of motion (\ref{omega:time_evo}) for the aspect ratio $\omega$ of particles, which couples to the equations for the nematic order parameter (\ref{EoM_Q}) and velocity fields (\ref{EoM_fluid}) through the flow alignment parameter $\xi$ and the nematic bulk free energy $f_{LC}$. 

By analysing the linear stability of isotropic systems with initially circular nematogens, we showed that they must  overcome an activity threshold in order to extend sufficiently to drive flows. Above this threshold the system enters a dynamical steady state characterised by coexisting regions of elongated particles, which tend to align nematically and are associated with topological defects and spatiotemporal chaotic flows, and quiescent regions consisting of isotropic particles. While isotropic particles get elongated by tissue-scale flow \ju{gradients} from neighbouring nematic regions, there is little extensional flow deep in the nematic regions, where the aspect ratio of particles subsequently decreases due to elastic shape energy. In steady-state, particles continuously get stretched by flow \ju{gradients} and subsequently contract in the absence of flow \ju{gradients}, leading to large variations of cell shape. Similar cell shape variations have been observed in simulations of the vertex model for unjammed tissues \cite{bi2016motility} and interpreted in terms of a continuum model.\cite{poissonbracket} Mean field theories of cell shape variations based on the vertex cellular Potts models have also been presented.\cite{sadhukhan2022origin, czajkowski2018hydrodynamics}

We have focused on extensile materials. Contractile systems which are initially circular will not show any dynamical behaviour because active forces just reinforce the thermodynamic forces restoring the particles to circular. \ju{We note, however, that our model is restricted to nematic driving: polar forces which can lead to coherent motion of circular cells are also observed in cell monolayers, and the resultant force on a cell does not necessarily act along its nematic axis. \cite{ladoux2017mechanobiology, ron2023polarization}}

Our model was motivated by the observation of active turbulence in confluent cell layers where cells can be deformed. In the light of our results it would be interesting to further investigate the spatial and temporal variation of the aspect ratio of cells within confluent layers that are on average circular, such as the MDCK cell line.\cite{balasubramaniam2021investigating, saw2017topological}



\section*{Author Contributions}
\ju{All authors worked on defining the project, analysing data and writing the paper. I. H. ran the simulations.}

\section*{Conflicts of interest}
There are no conflicts to declare.

\section*{Acknowledgements}
This project was funded by the Gould \& Watson Scholarship and the European Commission’s Horizon 2020 research and innovation programme under the Marie Sklodowska-Curie grant agreement No 812780.

\appendix

\section{Dispersion Relation Derivation}\label{append:C}
The inverse Fourier transform for a fluctuating field $f'$ is given by 
\begin{align}
f'(\mathbf{r}, t)= \int d\lambda dq \Tilde{f}(\mathbf{q}, \lambda)e^{i \mathbf{q}\cdot \mathbf{r} + \lambda t}.
\end{align}
The time evolution equation for the tensorial order parameter $\mathbf{Q}$ is eliminated by setting $\mathbf{n} = \hat{x}$ and $S = \left(\omega - 1\right)/\omega$. The relevant equations of motion to first order in the perturbed fields are then:
\begin{equation}\label{omega:perturb_append}
    \partial_{t}\omega' = 2E_{\parallel}' - \Gamma_{\omega}A_{\omega}\omega' + \dfrac{3}{2}\Gamma_{\omega}K_{LC}\nabla^{2}\omega',
\end{equation}
\\
\begin{equation}\label{vel:perturb_append}
    \rho \partial_{t}u_{i}' = \partial_{j}\left(2\eta E_{ij}' - p'\delta_{ij} - \dfrac{3}{2}\zeta \omega'(n_{i}n_{j} - \dfrac{1}{3}\delta_{ij})\right).
\end{equation}
In Fourier space, incompressibility reads
\begin{align}\label{incomp}
    q_{j}\Tilde{u}_{j} = 0 \Rightarrow
\Tilde{u}_{x} = -\dfrac{\sin{\theta}}{\cos{\theta}}\Tilde{u}_{y}.
\end{align}
\noindent In Fourier space, Eq.~\eqref{vel:perturb_append} reads:
\begin{align}
    \rho \lambda \Tilde{u}_{i} = \eta\left(-q_{i}\underbrace{q_{j}\Tilde{u}_{j}}_{\text{= 0}} - q^{2}\Tilde{u}_{i}\right) - q_{i}\Tilde{p} - \dfrac{3i}{2}\zeta \Tilde{\omega}q_{j}\left(n_{i}n_{j} - \dfrac{1}{3}\delta_{ij}\right).
\end{align}
Taking $\mathbf{n} = \left(1,0,0\right)$ and $\mathbf{q} = q\left(\cos{\theta}, \sin{\theta}, 0\right)$ where $\theta$ is the angle between $\mathbf{n}$ and $\mathbf{q}$ and considering $i = x$, we get
\begin{align}
    \rho \lambda  \Tilde{u}_{x} = -\eta q^{2}\Tilde{u}_{x} - iq\cos{\theta}\Tilde{p} - i\zeta \Tilde{\omega}q \cos{\theta},
\end{align}
from which we can solve for $\Tilde{p}$
\begin{align}\label{pressure}
    \Tilde{p} = \dfrac{i}{q\cos{\theta}}\left(\rho \lambda + \eta q^{2}\right)\Tilde{u}_{x} - \zeta \Tilde{\omega} = - \zeta \Tilde{\omega} - \dfrac{i \sin{\theta}}{q \cos^{2}{\theta}}(\rho \lambda + \eta q^{2})\Tilde{u}_{y},
\end{align}
where we have used incompressibility \eqref{incomp} in the final step. 
Now consider $i = y$
\begin{align}
    \rho \lambda \Tilde{u}_{y} = -\eta q^{2}\Tilde{u}_{y} - iq \sin{\theta}\Tilde{p} - \dfrac{3i}{2}\zeta \Tilde{\omega}\left(-\dfrac{1}{3}q \sin{\theta}\right).
\end{align}
Substituting the expression \eqref{pressure} for $\Tilde{p}$
\begin{align}
    \hspace*{-0.9cm}
    \left(\rho \lambda + \eta q^{2}\right)\Tilde{u}_{y} = -\dfrac{\sin^{2}{\theta}}{\cos^{2}{\theta}}\left(\rho \lambda + \eta q^{2}\right)\Tilde{u}_{y} + iq\sin{\theta}\zeta \Tilde{\omega} + \dfrac{i}{2}q\sin{\theta}\zeta\Tilde{\omega},
\end{align}
allows us to solve for $\Tilde{u}_{y}$
\begin{align}\label{vely}
    \Tilde{u}_{y} = \dfrac{3iq\cos^{2}{\theta}\sin{\theta}\zeta \Tilde{\omega}}{2\left(\rho \lambda + \eta q^2\right)}.
\end{align}
Next, we consider Eq.~\eqref{omega:perturb_append} in Fourier space:
\begin{align}
    \lambda \Tilde{\omega} = 2iq\cos{\theta}\Tilde{u}_{x} - \Gamma_{\omega}A_{\omega}\Tilde{\omega} - \dfrac{3}{2}\Gamma_{\omega}K_{LC}q^2\Tilde{\omega}.
\end{align}
Substituting for $\Tilde{u}_{x}$ by using \eqref{vely} for $\Tilde{u}_{y}$ and the incompressibility relation \eqref{incomp}, we finally get
\begin{align}
    \lambda \Tilde{\omega} = \dfrac{3q^{2}\cos^{2}{\theta}\sin^{2}{\theta}\zeta}{\left(\rho \lambda + \eta q^2\right)}\Tilde{\omega} - \Gamma_{\omega}A_{\omega}\Tilde{\omega} - \dfrac{3}{2}\Gamma_{\omega}K_{LC}q^2\Tilde{\omega},
\end{align}
which, after some algebraic manipulation and a double angle formula gives the dispersion relation
\begin{align}
    &q^{2}\left(\left(\eta +  \dfrac{3}{2}\Gamma_{\omega}K_{LC}\rho\right)\lambda + \Gamma_{\omega}A_{\omega}\eta - \dfrac{3}{4}\zeta \sin^{2}{2\theta}\right)\\ 
    &+ \dfrac{3}{2}\Gamma_{\omega}K_{LC}\eta q^{4} + \rho \lambda^{2} + \Gamma_{\omega}A_{\omega}\rho\lambda = 0.
\end{align}
For a solution with $\lambda > 0$ to exist for some value of $q$, we require the coefficient of $q^{2}$ to be negative. There is then a critical value of the activity, below which an isotropic system will be stable, given by
\begin{align}
    \zeta_{c} = \dfrac{4\Gamma_{\omega}A_{\omega}\eta}{3\sin^{2}{2\theta}}.
\end{align}
$\zeta_{c}$ is clearly minimised when $\theta = \pi/4$. 
An analytical expression may be obtained for the most unstable wave vector, $q_{m}$ which corresponds to the local maximum of the growth rate, $\lambda_{m}$. This is given by:
\begin{align}
    &q_{m}^{2} = \dfrac{\sqrt{2\rho K_{LC}\Gamma_{\omega}\zeta \sin^{2}{2\theta} J_{+}^{2}R} - 4\rho K_{LC}\Gamma_{\omega}\eta T}{2K_{LC}\Gamma_{\omega}\eta J_{-}^{2}},\\
    &J_{\pm} = 2\eta \pm 3\rho\Gamma_{\omega}K_{LC},\\
    &R = 3\zeta\sin^{2}{2\theta} - 4 \Gamma_{\omega}A_{\omega}\eta + 6\rho A_{\omega}K_{LC}\Gamma_{\omega}^{2},\\
    &T = 3\zeta \sin^{2}{2\theta} - \Gamma_{\omega}A_{\omega}J_{-}.
\end{align}
Despite the complicated form of the above expression, the competition between activity versus elasticity and energy is shown to be a determining factor for the most unstable modes. 
\\
Finally, by setting $\omega' = \epsilon \sin{\mathbf{q}\cdot \mathbf{r}}$ and using the Fourier transform of the $\sin$ function, we may also find a solution of the flow field, $\Tilde{u}_{y}$.
\begin{align}
    \Tilde{u}_{y} = \dfrac{3iq\cos^{2}{\theta}\sin{\theta}\zeta}{2(\rho \lambda + \eta q^2)}\dfrac{1}{2i} (2\pi)^{2}\epsilon \left(\delta(\mathbf{q} + \mathbf{q'}) - \delta(\mathbf{q} - \mathbf{q'})\right).
\end{align}
Setting $\lambda = 0$ for simplicity, $\theta = \pi/4$, and inverting the Fourier transform, we may get the full flow field:
\begin{align}
    &u_{y} = -\dfrac{3}{4\sqrt{2}}\dfrac{\zeta \epsilon}{\eta q}\cos\left(\dfrac{1}{\sqrt{2}}(x + y)\right),\\
    &u_{x} = \dfrac{3}{4\sqrt{2}}\dfrac{\zeta \epsilon}{\eta q}\cos\left(\dfrac{1}{\sqrt{2}}(x + y)\right).
\end{align}


\balance


\bibliography{rsc} 
\bibliographystyle{rsc} 
\end{document}